\documentclass[a4paper,twoside]{article}

\usepackage{epsfig}
\usepackage{calc}
\usepackage{amssymb}
\usepackage{amstext}
\usepackage{amsmath,lipsum}
\usepackage{amsthm}
\usepackage{multicol}
\usepackage{pslatex}
\usepackage{apalike}
\usepackage{hyperref}
\usepackage{booktabs}
\usepackage{tabularx}
\usepackage{graphicx}
\usepackage{fontawesome5}
\usepackage{threeparttable}
\usepackage{subfig}
\theoremstyle{definition}
\newtheorem{definition}{Definition}[section]
\usepackage{algorithm2e}
\usepackage[bottom]{footmisc}
\usepackage{SCITEPRESS}  

\begin{document}

\title{Knowledge Discovery in Optical Music Recognition: Enhancing Information Retrieval with Instance Segmentation}

\author{\authorname{Elona Shatri\sup{1}\orcidAuthor{0000-0002-1651-5848}, Gy\"orgy Fazekas\sup{1}\orcidAuthor{0000-0003-2580-0007} }
\affiliation{\sup{1}Queen Mary University of London, London, United Kingdom}
\email{\{e.shatri, g.fazekas\}@qmul.ac.uk}
}


\abstract{Optical Music Recognition (OMR) automates the transcription of musical notation from images into machine-readable formats like MusicXML, MEI, or MIDI, significantly reducing the costs and time of manual transcription. This study explores knowledge discovery in OMR by applying instance segmentation using Mask R-CNN to enhance the detection and delineation of musical symbols in sheet music. Unlike Optical Character Recognition (OCR), OMR must handle the intricate semantics of Common Western Music Notation (CWMN), where symbol meanings depend on shape, position, and context. Our approach leverages instance segmentation to manage the density and overlap of musical symbols, facilitating more precise information retrieval from music scores. Evaluations on the DoReMi and MUSCIMA++ datasets demonstrate substantial improvements, with our method achieving a mean Average Precision (mAP) of up to 59.70\% in dense symbol environments, achieving comparable results to object detection. Furthermore, using traditional computer vision techniques, we add a parallel step for staff detection to infer the pitch for the recognised symbols. This study emphasises the role of pixel-wise segmentation in advancing accurate music symbol recognition, contributing to knowledge discovery in OMR. Our findings indicate that instance segmentation provides more precise representations of musical symbols, particularly in densely populated scores, advancing OMR technology. We make our implementation, pre-processing scripts, trained models, and evaluation results publicly available to support further research and development.}

\onecolumn \maketitle \normalsize \setcounter{footnote}{0} \vfill
\section{\uppercase{Introduction}}
\label{sec:intro}

Optical Music Recognition (OMR) is a research subfield within Music Information Retrieval that automates the transcription of music notation from images into machine-readable formats such as MusicXML~\cite{b70}, MEI~\cite{b71}, or MIDI\footnote{https://www.midi.org}. This automation addresses the significant time and cost involved in manually transcribing music scores, a process essential for digital music analysis, editing, and playback. Beyond automation, OMR has profound implications for knowledge discovery and information retrieval within large music databases. By converting analogue musical scores into searchable digital formats, OMR enables researchers, educators, and musicians to efficiently explore vast collections of musical works, facilitating musicological research, comparative studies, and the discovery of patterns and trends across different musical eras and styles. These capabilities underscore OMR's critical role in advancing music information retrieval and digital humanities.

While OMR is often compared to Optical Character Recognition (OCR), which transcribes printed text into digital form, OMR faces unique complexities due to the multifaceted nature of Common Western Music Notation (CWMN). In music scores, the meaning of symbols depends not only on their shapes but also on their precise positions on the staff and their contextual relationships with other symbols. For example, a note's value and pitch are determined by its position on the staff and its interaction with key signatures, time signatures, and other musical elements. Consequently, accurate staff detection is crucial, so our approach incorporates a parallel step using traditional computer vision techniques to address this challenge, as detailed in Subsection \ref{subsec:detection-staff}. These intricacies present challenges that standard OCR methods, primarily designed for straightforward text recognition, cannot handle. Therefore, specialised techniques are required to interpret and digitise musical notation accurately.


Recent advances in deep learning, particularly in the use of Convolutional Neural Networks (CNN) \cite{b10,pacha2018optical}, Recurrent Neural Networks (RNN) \cite{baro2018optical}, and more recently, transformer-based models \cite{li2023tromr,rios2023sheet,rios2024sheet}, have significantly enhanced the capabilities of OMR systems. CNNs excel at recognising patterns and features within images, making them well-suited for identifying and distinguishing musical symbols. RNNs, in contrast, are adept at handling sequential data, enabling them to interpret the temporal and contextual relationships between musical elements. Together, these technologies have improved the accuracy of symbol detection and interpretation in OMR. Despite these advances, significant limitations remain, particularly when dealing with densely packed and overlapping symbols, which can result in errors in detection and interpretation \cite{pacha2017towards}.

Our study compares object detection with instance segmentation using Mask R-CNN \cite{b42}, integrating a parallel staff detection stage for pitch inference. This step addresses the challenge of recognising thin, elongated staff lines, which detection methods often struggle with. Instance segmentation enables pixel-level classification to precisely delineate musical symbols, particularly in dense and overlapping notation, enhancing OMR accuracy and supporting reliable information extraction from music scores.

We also emphasise the importance of significantly improving performance by domain-specific pre-training, such as using MUSCIMA++ weights. Our focus on full-page music symbol recognition addresses the complexities of entire music sheets, distinguishing our approach from methods that only handle isolated symbols.

By improving musical symbol recognition, our method facilitates knowledge discovery and information retrieval in large music databases through enhanced metadata extraction, enabling advanced queries and comparative musicology. We evaluate our approach using the DoReMi and MUSCIMA++ datasets and provide our implementation details here \href{https://github.com/elonashatri/pitch_Mask_RCNN.git}{here} \footnote{https://github.com/elonashatri/pitch\_mask\_rcnn}.


Our study shows that using models like Mask R-CNN for instance segmentation, along with efficient staff detection algorithms, advances OMR for full-page images. These models are less data-hungry than transformer-based approaches, making them more practical for OMR, where annotated datasets are limited. This combination provides a more accurate and efficient solution for digitising musical scores, supporting enhanced music analysis, retrieval, and knowledge discovery.

\section{\uppercase{Related Work}}\label{sec:background}

OMR has traditionally been approached through four stages: image pre-processing, musical object detection, reconstruction, and encoding \cite{b8,b24}. The primary objective of musical symbol detection is to find the bounding boxes and corresponding classes of musical objects. Undetected musical primitives in this stage can introduce errors into subsequent stages, making detection a critical component of OMR. This is particularly important due to the complexities introduced by artefacts, image quality, and object density. The classified primitive elements are then integrated using graphical or syntactic rules, or more recently, deep learning, to reconstruct the musical notation in the third stage. The final encoding stage converts the reconstructed notation into a machine-readable format mentioned in Section \ref{sec:intro}. 


Traditional object detection in OMR has seen the application of models like Region-based CNNs (R-CNNs) \cite{b65}, which integrate region proposals with CNNs. Despite their effectiveness in certain domains, these models face limitations in OMR due to their separate training stages and the challenges of handling densely packed musical symbols. Fast R-CNN and Faster R-CNN \cite{b48,b45} addressed some of these issues by introducing a Region Proposal Network (RPN) that generates region proposals more efficiently, sharing convolution features with the detection network. However, while effectively reducing computational costs and improving detection accuracy, these models still struggle with full-page musical scores and densely populated notation \cite{pacha2018optical}.

Beyond object detection, semantic segmentation has been explored to provide pixel-level classification of similar entities within images \cite{b4,b38}. Techniques like the Deep Watershed Detector \cite{b55} and U-Net architectures \cite{b49} have been employed to improve the detection of smaller musical objects and overlapping symbols. However, these methods also have limitations. The Deep Watershed Detector struggles with larger musical objects and rare classes due to its size variation learning limitations. At the same time, U-Net models are sensitive to training data distribution and may lose translation equivalence due to their down-sampling and up-sampling processes \cite{b2,b43}.
Instance segmentation models, such as Mask R-CNN \cite{b42}, offers a more refined approach by providing pixel-level classification for each object, allowing for the precise identification and separation of overlapping symbols. This method extends Faster R-CNN by adding a branch that predicts segmentation masks on each Region of Interest (RoI), using RoI Align to avoid the quantisation issues seen in earlier models. By generating pixel-wise masks, instance segmentation improves the detection and analysis of musical symbols and facilitates the exclusion of noise and more accurate pitch identification relative to staff lines. Mask R-CNN has been applied to handwritten 4-part harmonies \cite{de2022optical} showing promising results in these types of scores.

Despite the advancements brought by instance segmentation, challenges remain, particularly with class imbalance in musical notation datasets and the variability in handwritten music. Certain symbols, such as noteheads and stems, are more prevalent, skewing the training process. Additionally, detecting thin, closely spaced objects like staff lines and barlines poses significant difficulties.

Our study addresses these challenges by implementing and comparing object detection and instance segmentation techniques for OMR using the DoReMi and MUSCIMA++ datasets. We demonstrate the effectiveness of Mask R-CNN in handling dense and overlapping symbols and provide a comprehensive comparative evaluation highlighting the advantages of instance segmentation over traditional methods. By integrating a parallel staff detection stage, our approach enhances the accuracy and detail of OMR systems and offers a more comprehensive solution for digitising musical notation. Implementation details and evaluation results will be publicly available.


\begin{figure*}
\centerline{
\includegraphics[width=1.8\columnwidth]{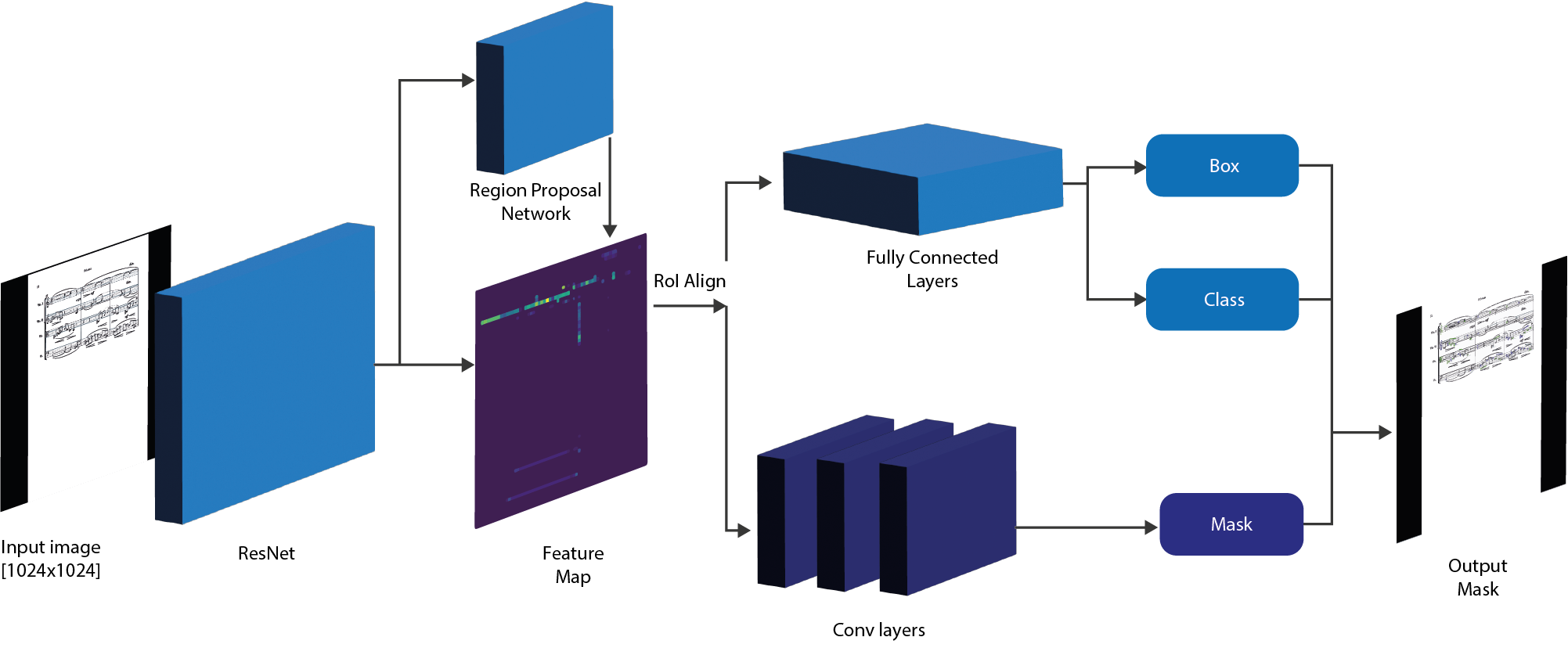}}
\caption{ Mask R-CNN architectures applied to OMR}
\label{fig:net-architecture}
\end{figure*}


    


\section{\uppercase{Datasets}}
This study utilises two prominent datasets in the domain of OMR: MUSCIMA++ \cite{b44} and DoReMi \cite{b28}. Both datasets play crucial roles in training and evaluating our models, offering diverse challenges and benefits.

\paragraph{MUSCIMA++} The MUSCIMA++ dataset consists of handwritten musical scores and is specifically designed to address the challenges of handwritten OMR. It provides a substantial number of annotated images with detailed metadata files that include bounding boxes and pixel masks for each musical symbol. This dataset is particularly valuable for its representation of handwritten music, which introduces variability and complexity not found in typeset scores.

\paragraph{DoReMi} The DoReMi dataset comprises approximately 6,400 high-resolution images (300 DPI, 2475x3504 pixels) of typeset sheet music, annotated with one of 94 category labels. This dataset is notably larger than MUSCIMA++, making it a robust resource for training deep learning models. The high resolution and detailed annotations allow for precise training and evaluation of models aimed at both object detection and instance segmentation.
Both datasets exhibit class imbalance, with a significant portion of the annotated objects being stems and noteheads. This imbalance presents a challenge for training models that need to recognise a diverse set of musical symbols.

The DoReMi dataset is primarily used for training our models due to its larger size and the fact that it consists of typeset images, which are more standardised than handwritten scores. 
In our object detection experiments, we train different models using varying subsets of the DoReMi dataset: 27\%, 45\%, 90\%, and 100\% of the data. This stratified approach serves two key purposes:

\begin{itemize}
    \item It helps identify the most effective amount of data required for training without overfitting.
    \item Explore how class balance affects the detection rate by avoiding the predominance of infrequent classes that could skew the training process.
\end{itemize}

The resulting datasets used for training consist of 94, 71, 71, and 64 classes, respectively. For instance segmentation, we further refine our subsets to include 291 and 1685 images from DoReMi, focusing on limiting computational time during training while maintaining a sufficient number of classes to evaluate model performance effectively. By iteratively refining the training process and expanding the dataset, this study aims to identify key factors that may improve the accuracy and reliability of segmenting musical symbols in sheet music.
\section{\uppercase{Methodology}}\label{sec:method}
This section discusses the detailed methodology, covering object detection, instance segmentation, data pre-processing, and training configurations.
Our approach leverages full-page images, encompassing more objects with smaller bounding boxes than staff-cropped images. This approach enhances the granularity and accuracy of detection and segmentation within the full-page images.

We conduct two experiments using the DoReMi dataset: one on detecting music notation primitives and another on instance segmentation. Object detection predicts a bounding box $(x_1, x_2, y_1, y_2)$, an associated category, and a confidence score for each musical element, such as stems, noteheads, or dynamics (e.g., {\em ppp}). Both pipelines include a parallel step for staff detection using a traditional method described in Section \ref{subsec:detection-staff}.


\subsection{Music Symbol Object Detection}
For the object detection task, we employ Faster R-CNN, a well-established model in object detection. Faster R-CNN integrates a Region Proposal Network (RPN) with a Fast R-CNN detector, allowing for efficient and accurate object localisation and classification. The loss function for Faster R-CNN is defined as:

\begin{equation}
\begin{aligned}
L\left(\left\{p_{i}\right\},\left\{t_{i}\right\}\right)  &= \frac{1}{N_{c l s}} \sum_{i} L_{c l s}\left(p_{i}, p_{i}^{*}\right) \\
&+  \lambda \frac{1}{N_{\text {reg }}} \sum_{i} p_{i}^{*} L_{r e g}\left(t_{i}, t_{i}^{*}\right),
\end{aligned}
\label{eq:faster-rcnn}
\end{equation}

where $i$ is the index of an anchor in a mini-batch and $p_{i}$ is the predicted probability of anchor $i$ being an object. We use two losses, classification loss $L_{cls}$ and regression loss $L_{reg}$. Classification loss is log loss over two classes (object vs not object). Regression loss is activated only for positive anchors $p_{i}^{*} L_{reg}$ and not activated otherwise. Both terms are then normalised by $N_{cls}$ and $N_{reg}$ and weighted by $\lambda$ as a balancing parameter. Every bounding-box has the associated set of cells in the feature map computed by a CNN. Our Faster R-CNN implementation follows prior work from \cite{b45,b10}.

The Faster R-CNN model configurations employ ResNet50 or Inception ResNet v2 \cite{b30} backbones with Atrous convolutions, optimised for the MUSCIMA++ or COCO \cite{b39} datasets with image dimensions of 2475×3504 pixels. The model maintains aspect ratios, with image dimensions between 500 and 1000 pixels. The feature extractor uses a stride of 8 in the first stage, and anchor generation is handled by a grid anchor generator with various widths, heights, scales, and aspect ratios. The first-stage Atrous rate is 2, with box predictor parameters using an L2 regulariser and a truncated normal initialiser. Non-maximum suppression (NMS) has an IoU threshold of 0.5 with up to 500 proposals. The initial crop size is 17, and max-pooling has a kernel size and stride of 1. The model is trained with a batch size of 1 using an RMSProp optimiser (initial learning rate of 0.003, decaying by 0.95 every 30,000 steps), with momentum at 0.9 and gradient clipping at 10.0. Data augmentation includes random horizontal flipping, and evaluation metrics follow the COCO standard, using 120 examples for evaluation.


\subsection{Music Symbol Instance Segmentation}

For instance segmentation, we utilise Mask R-CNN \cite{b42}, which extends Faster R-CNN by adding a branch for predicting segmentation masks on each Region of Interest (RoI), as shown in Figure \ref{fig:net-architecture}). This method allows for pixel-level classification of objects, which is crucial for accurately delineating overlapping musical symbols. We define the task of instance segmentation as follows:

\begin{definition} Music Symbol Instance Segmentation (MSIS) is the task of assigning a music symbol class to each pixel of a sheet music image in addition to predicting bounding boxes with their corresponding confidence scores.
\end{definition}
The loss function in instance segmentation is given as:

\begin{equation}
\begin{aligned}
Total Loss ={} & rpn\_class\_loss + rpn\_bbox\_loss \\
                      & + mrcnn\_class\_loss + mrcnn\_bbox\_loss \\
                      & + mrcnn\_mask\_loss
\end{aligned}
\label{eq:mask-rcnn}
\end{equation}

Here, the total loss is a combination of several components: RPN Class Loss, which measures the error in classifying the anchors; RPN Bounding Box Loss, which quantifies the error in bounding box predictions by the RPN; Mask R-CNN Class Loss, which evaluates the error in classifying objects within the proposed regions (RoIs); Mask R-CNN Bounding Box Loss, which measures the error in bounding box predictions for the RoIs; and Mask R-CNN Mask Loss, which evaluates the accuracy of the predicted segmentation masks for each RoI. These components collectively ensure accurate detection and segmentation of musical symbols in sheet music images.

We utilise two backbone networks, ResNet50 and ResNet101 \cite{b32}, which are pre-trained either on the COCO dataset or a subset of the DoReMi dataset. The training process is structured in multiple phases to enhance model performance iteratively. Initially, the model is trained on a dataset of 291 images with weights pre-trained on COCO, focusing only on the head layers for 20 epochs. Subsequently, training is extended to all layers for an additional ten epochs using the same dataset. To further improve performance, the model is trained on an expanded dataset of 1,348 images for another 40 epochs, fine-tuning the weights from the previous phase.

For the best model, the ResNet101 backbone network has strides of [4, 8, 16, 32, 64] and a batch size of 1. Detection parameters included a maximum of 100 instances, a minimum confidence of 0.9, and an NMS threshold of 0.3. The FPN classification fully connected layer size is 1024. The gradient clipping norm is set to 5.0, and we used one image per GPU. The image dimensions are resized to a square shape of [1024, 1024, 3], and the mini mask shape is (56, 56). The learning momentum is 0.9, with a learning rate of 0.0001. The dataset comprised 72 classes. The RoI positive ratio is 0.33, and RPN anchor ratios were [0.5, 1, 2] with scales (32, 64, 128, 256, 512) and stride 1. RPN bounding box standard deviation was [0.1, 0.1, 0.2, 0.2] with an NMS threshold of 0.7 and 256 RPN train anchors per image.
Training is re-initiated from epoch 39, using pre-trained weights from the specified path with a learning rate of 0.0001, and the network was trained on all layers.

\begin{table*}[t]
\small
  \centering
  \begin{tabularx}{\textwidth}{>{\centering\arraybackslash}X@{\hspace{63pt}}>{\centering\arraybackslash}X@{\hspace{10pt}}>{\centering\arraybackslash}X@{\hspace{2pt}}>{\centering\arraybackslash}X@{\hspace{15pt}}>{\centering\arraybackslash}X@{\hspace{10pt}}>{\centering\arraybackslash}X@{\hspace{10pt}}>{\centering\arraybackslash}X@{\hspace{10pt}}>{\centering\arraybackslash}X@{\hspace{6pt}}>{\centering\arraybackslash}X}
    \hline
        Model & Classes & Data (\%) &  Steps  & Large  & Medium  & Small & mAP \\ \hline
        
        InceptionResNetV2* & 94  & 100 & 120K & 32.53 & 38.24 & 15.46 & 47.37 \\
        InceptionResNetV2* &  71  & 90 & 120K & 36.79 & 47.42 & 25.55 & 63.45 \\
        
        InceptionResNetV2*  & 71  & 45 & 120K & 33.89 & 54.08 & 22.38 & 65.42 \\
        
        InceptionResNetV2*  &  67  & 27 & 120K & 40.97 & 46.45 & 27.13 & 63.70 \\
        
        InceptionResNetV2+  & 71 & 90 & 120K & 41.08 & 49.93 & 23.18 & 64.92\\
        
        ResNet50* & 71 & 90 & 120K & 36.73 & 45.22 & 28.98 & 59.81\\
        
        ResNet50*  & 71  & 45 & 120K & 32.79 & 46.03 & 31.76 & 62.45\\
        
        ResNet50*  &  67  & 27 & 120K & 39.32 &  46.64 &  29.62 & 63.19\\
        
        ResNet50+ &  71 & 90 & 120K & 35.90 & 44.57 &  29.17 & 63.13\\

        ResNet50* & 94 & 100 & 80K & \textbf{93.63}  & \textbf{75.82} & \textbf{41.71} & \textbf{80.99}\\
    \hline
        InceptionResNetV2† & 63 & CS63 & - & \textbf{-}  & \textbf{-} & \textbf{-} & \textbf{64.1}\\
        CascadeRCNN†  & 136 & DS & - & \textbf{-}  & \textbf{-} & \textbf{-} & \textbf{70.0}\\
    \hline\\
  \end{tabularx}
  \caption{Object detection results using DoReMi dataset, '*' denotes COCO weights used for initialisation and '+' denotes MUSCIMA++ weights. Mean Average Precision (mAP) given at 0.50 IoU (\%). The Large, Medium and Small columns show the mAP for large, medium and small objects respectively. The last two rows present experimental results using InceptionResNetV2 in CollabScore63, a variant of DeepScores with a reduced number of classes, utilising a training set of 1,362 images, alongside their proposed Cascade R-CNN - FocalNet model with 136 classes \cite{yesilkanat2023full}.}
  \label{table:baseline-results-faster-r-cnn-inception}
\end{table*}

\begin{table*}
\centering
\begin{threeparttable}

    \begin{tabular}{lcccccc}
    \hline
    Feature Extractor & Weights & No. of Images & Epochs & Layers & mAP@.50\% \\
    \hline
    ResNet50 & COCO & 291 & 20 & Heads & 16.239 \\
    ResNet101 & COCO & 291 & 30 & Heads & 37.151 \\
    ResNet101 & DoReMi & 291 & 30 + 10 & All & 46.087 \\
    \textbf{ResNet101} & \textbf{DoReMi} & 1384 & \textbf{80} & \textbf{All} & \textbf{59.70} \\
    \hline
    \end{tabular}
    
\end{threeparttable}
\caption{Performance metrics of various configurations of the ResNet model using the Mask R-CNN architecture.}
\label{table:mask-rcnn-results}
\end{table*}

The experimental setup and results are detailed in Table \ref{table:mask-rcnn-results}. Initially, the model training started on a relatively small dataset comprising 291 images, focusing exclusively on the network's head layer, employing pre-trained weights for initialisation (detailed in Table \ref{table:mask-rcnn-results}, 3rd row). Two subsequent stages of fine-tuning followed this phase:

\begin{itemize}
    \item After the initial focus on the head layer, training was extended to all network layers for an additional ten epochs. This stage aimed to refine the feature extraction capabilities across the entire network.
    \item To further examine the effects of dataset size on segmentation performance, the model underwent additional training on an expanded dataset of 1,348 images spanning 50 epochs. This stage evaluated how larger datasets influence the model's ability to generalise and improve segmentation accuracy.
\end{itemize}

This iterative training approach is structured to assess the effectiveness of pre-trained models and explore the potential of increasing dataset sizes to enhance music symbol segmentation accuracy, a notable challenge in OMR. 

\subsection{Staff Detection}\label{subsec:detection-staff}

The proposed methodology for detecting staff lines in musical notation involves several key steps. Initially, the input image is converted to grayscale and enhanced using Otsu's thresholding method to create a binary image. Horizontal lines are then detected through morphological operations. Contours representing potential staff lines are identified and analysed based on their geometric properties. Specifically, contours with a high aspect ratio and appropriate height are classified as staff lines, while others are marked as non-staff lines.

The large contours are divided into smaller segments to manage those that span multiple staff lines. The identified staff lines are then drawn on the image for visualisation (see Figure \ref{fig:full-mask} and \ref{fig:full-faster}), using different colours to differentiate between staff lines (in green) and non-staff lines (in red). The final processed image is displayed alongside the original for clear comparison in Figures \ref{fig:full-mask} and \ref{fig:full-faster}. This method aims to detect staff lines in musical notation, making it easier to output a more structured result by stave and enabling pitch or staffing position information to be captured. However, it is important to note that this approach may not work well with images of low quality. Still, it is robust enough to handle small perspective shifts, rotations, changes in contrast, and similar variations. Another limitation is that this method is based on the relative distance of the objects to the detected staff lines. The method may provide the wrong stave for complex scores, where musical objects such as noteheads are closer to the next stave than their own staff. This can be addressed using CNN trained to distinguish staves \cite{b76}.

This method has been applied in parallel with both instance segmentation and object detection models since they struggle with detecting thin, long objects such as staff lines.

\section{\uppercase{Evaluation and Results}}\label{sec:eval}

  

\begin{figure*}[ht]
  \centering
  \subfloat[Output from the Mask R-CNN model, each class is represented by a different colour]{\includegraphics[width=0.47\textwidth]{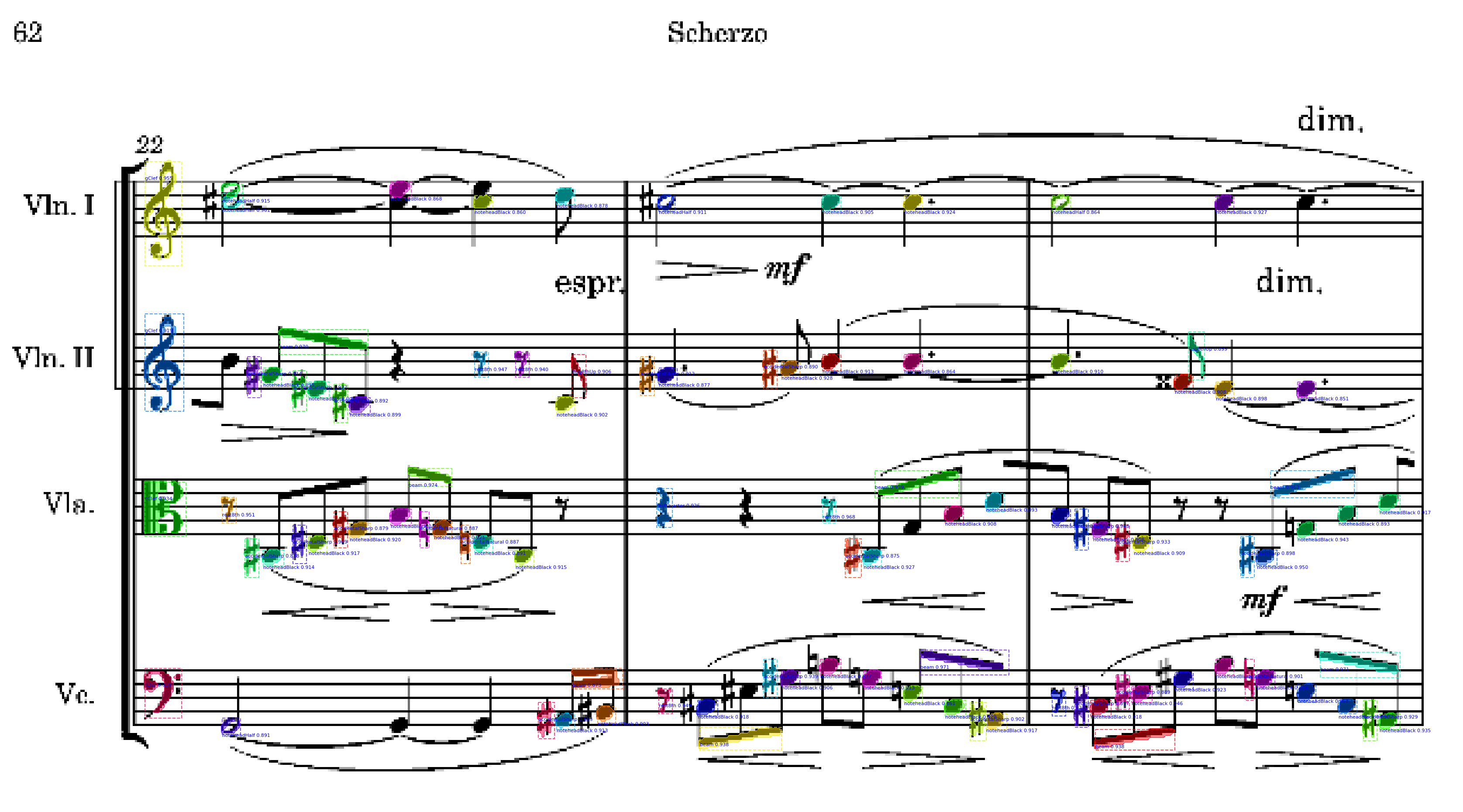}\label{fig:mask-output}}
  \hfill
  \subfloat[Concatenated Mask R-CNN output and staff detection, the green lines represent the detected staff lines]{\includegraphics[width=0.47\textwidth]{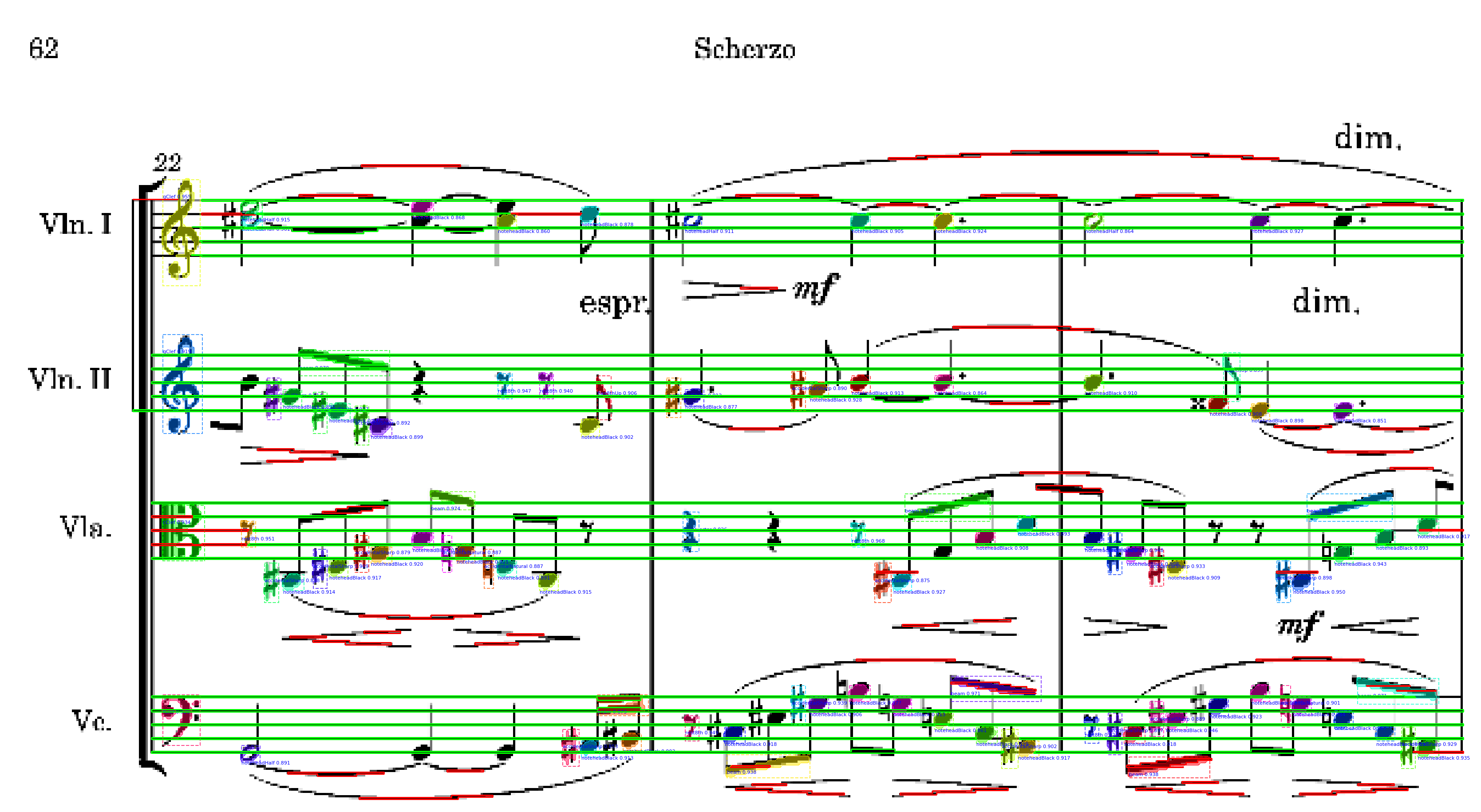}\label{fig:mask-output-staff}}
  \caption{Mask R-CNN model inference in an image}
  \label{fig:full-mask}
\end{figure*}


\begin{figure*}[h!]
  \centering
  \subfloat[Output from the Faster R-CNN model, each class is represented by a different colour]{\includegraphics[width=0.48\textwidth]{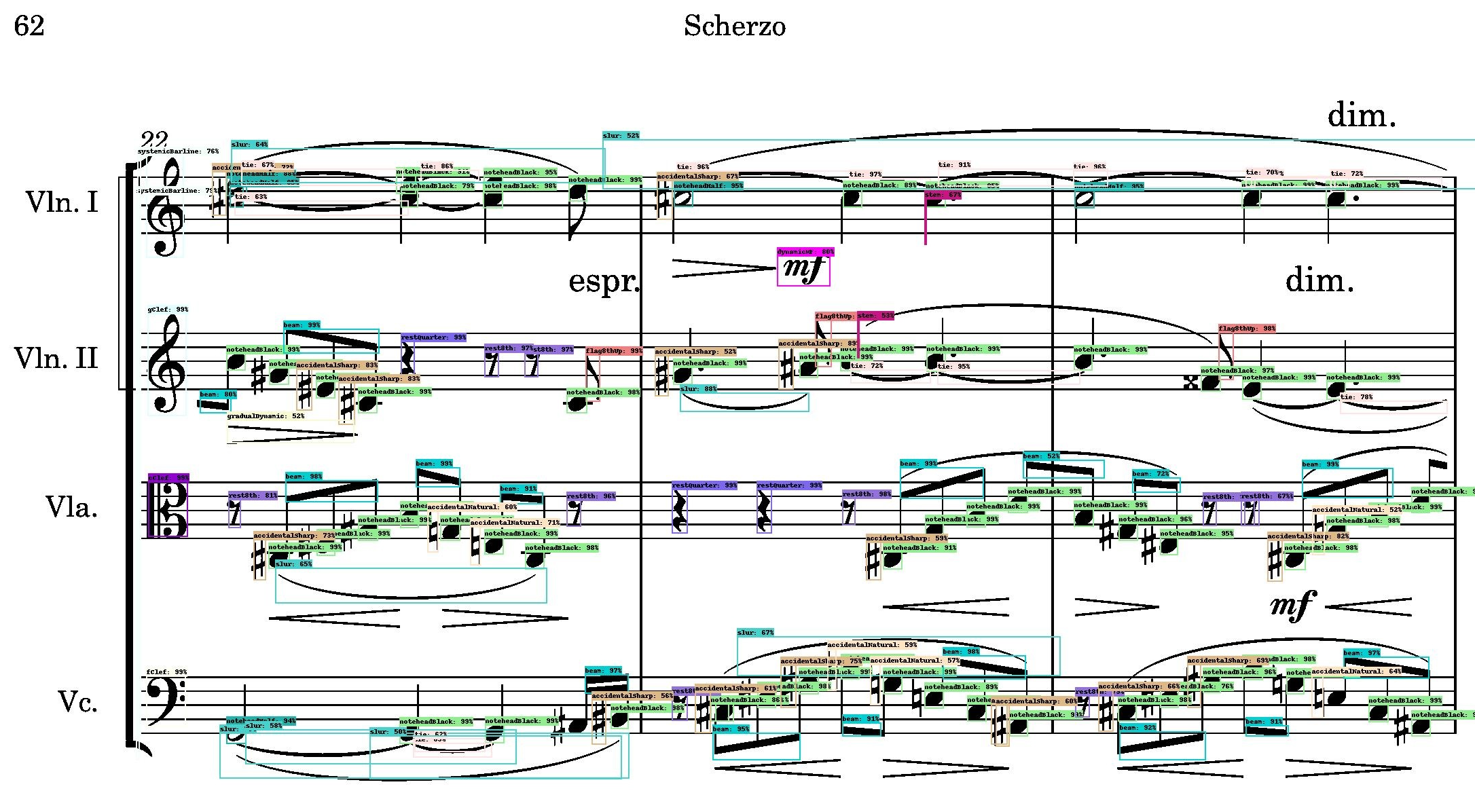}\label{fig:faster-output}}
  \hfill
  \subfloat[Concatenated Faster R-CNN output and staff detection, the green lines represent the detected staff lines]{\includegraphics[width=0.48\textwidth]{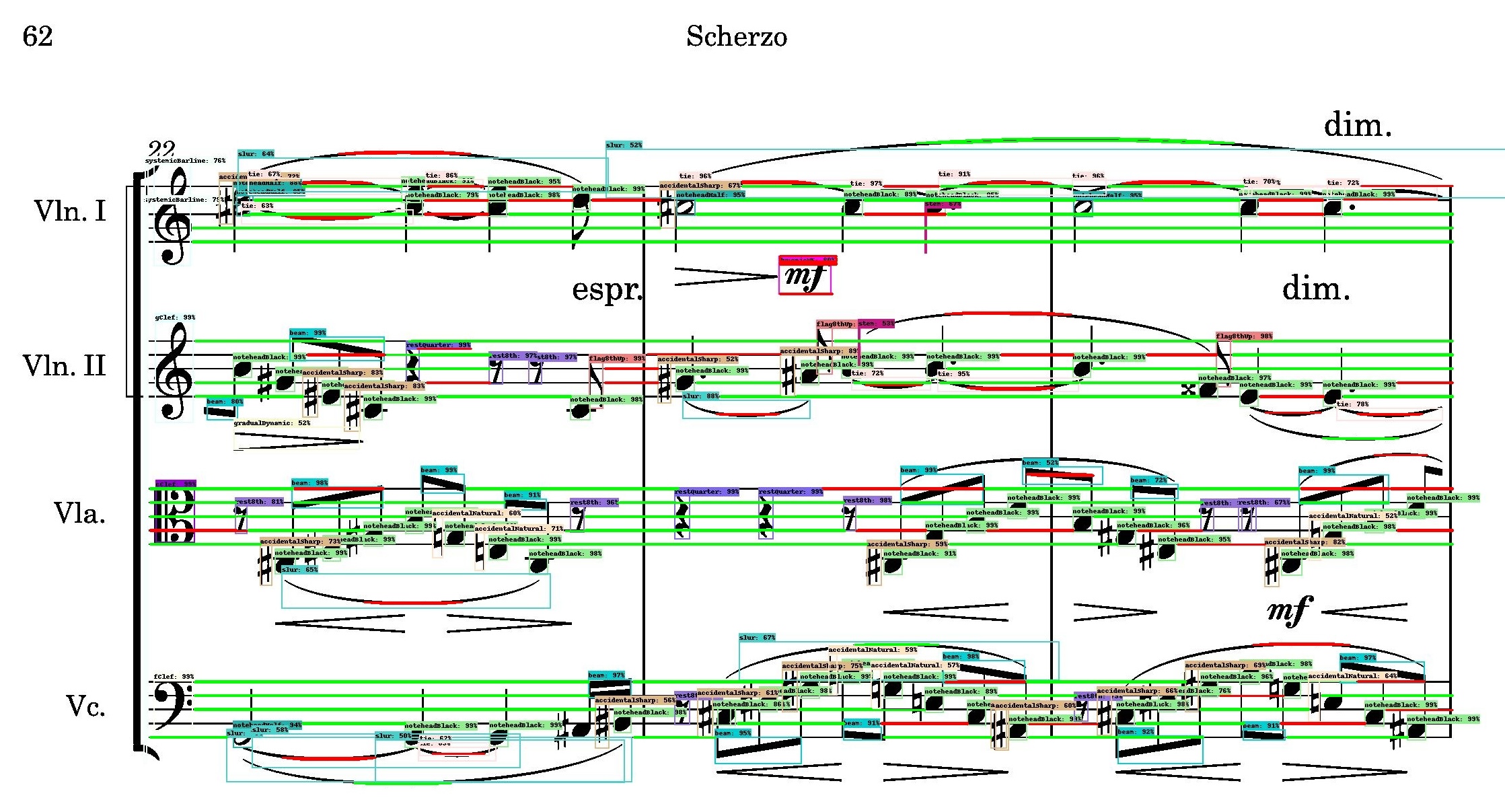}\label{fig:faster-output-staff}}
  \caption{Faster R-CNN model inference in an image}
  \label{fig:full-faster}
\end{figure*}

A key contribution of this study is the comparative analysis of object detection and instance segmentation methods for OMR. 
By evaluating both approaches, we aim to determine which method is better suited to the complexities of musical notation, particularly in handling dense and overlapping symbols. Object detection offers efficient symbol localisation but lacks the pixel-level precision provided by instance segmentation. Our comparison highlights the strengths and weaknesses of each method, giving valuable insights into their applicability in different OMR scenarios.

In this section, we detail the experiments conducted to evaluate the performance of the instance segmentation approach using Mask R-CNN, compared to traditional object detection methods. We used the DoReMi and MUSCIMA++ datasets, which offer diverse challenges due to their handwritten and typeset musical scores, respectively.

The performance of our object detection and instance segmentation models was assessed using the mAP as is standard in the field \cite{b39}. The mAP scores were computed by setting a threshold for Intersection over Union (IoU) to evaluate the accuracy of the predicted bounding boxes against the ground truth.

The trade-off between precision and recall is particularly significant in the domain of OMR, where precise localisation of overlapping musical symbols is crucial. Precision measures the accuracy of the predictions, whereas recall assesses the model's ability to detect all relevant instances. These metrics form the basis of our evaluation using the Average Precision (AP) for each class at an IoU threshold of 0.50.

\paragraph{Musical Symbol Object Detection}
We first applied Faster R-CNN with both InceptionResNetV2 and ResNet50 backbones for object detection tasks. The evaluation metrics were based on the mean Average Precision (mAP) at an Intersection over Union (IoU) threshold of 0.50. The results are summarised in Table 1. Using the InceptionResNetV2 backbone, we found that reducing the number of classes from 94 to 71 significantly improved mAP, with a peak mAP of 65.42\% achieved with 45\% of the data. Interestingly, using 90\% of the data resulted in a slightly lower mAP of 64.92\%, suggesting that a well-balanced and smaller dataset can sometimes outperform a larger one. Moreover, using domain-specific pre-trained weights from MUSCIMA++ provided nearly a 5\% improvement over COCO weights, underscoring the importance of domain adaptation.

Similarly, the ResNet50 backbone demonstrated notable performance, achieving a peak mAP of 63.19\% using only 27\% of the data. This result highlights the potential efficiency of a well-curated dataset. MUSCIMA++ weights also improved performance, although not as significantly as InceptionResNetV2. These results suggest that while both models are compelling, InceptionResNetV2 outperforms ResNet50, particularly with smaller training sets. However, this dynamic shifts when a larger dataset is used, with ResNet50* achieving an mAP of 80.99\% on the full dataset. This is comparable to similar models in OMR \cite{yesilkanat2023full}, which, in contrast, uses high-resolution input images, shown in the last two rows on Table \ref{table:baseline-results-faster-r-cnn-inception}. Finally, an investigation into average precision per category revealed that the model struggles to detect less frequent objects, such as rarely used time signatures, dynamics markers, and accidentals, as evident in the inference results shown in Figure \ref{fig:full-faster}.

\paragraph{Musical Symbol Instance Segmentation}

For instance segmentation, we employed Mask R-CNN with ResNet50 and ResNet101 backbones and evaluated using the Pascal VOC metric \cite{b75}. 
The results, detailed in Table \ref{table:mask-rcnn-results}, demonstrate significant performance improvements. Initially, using the ResNet50 backbone and training on 291 images with COCO pre-trained weights, the model achieved an mAP of 16.239\%. Further fine-tuning on the DoReMi dataset for 30 epochs increased the mAP to 46.087\%. In contrast, the ResNet101 backbone, when trained on the same 291 images, reached an mAP of 37.151\%. Extending the training to 1,384 images for 80 epochs with ResNet101 achieved the highest mAP of 59.70\%, demonstrating the benefits of a more extensive and more comprehensive dataset. These results are similar to the mAP achieved by object detection models shown in Table \ref{table:baseline-results-faster-r-cnn-inception} using a similar training set size. Findings from this evaluation indicate that:

\begin{itemize}
    \item Switching to a more complex network architecture (ResNet101) and training all layers comprehensively significantly improved mAP scores, confirming the architecture’s influence on performance.

    \item Extending training duration and increasing dataset size were crucial for achieving higher mAP scores. This is especially evident in the performance of ResNet101 when trained for 80 epochs on 1,384 images.

    \item The model faced difficulties predicting thin objects such as staff lines and stems, likely due to downsampling issues. Adjustments in RPN anchor ratios or pre-segmentation strategies might mitigate these challenges.
\end{itemize}


\paragraph{Beyond Detection and Segmentation}\label{para:staff-detection}

As stated in Section \ref{sec:intro}, the ultimate aim of OMR is to convert music scores into a structured, machine-readable file. While object detection and instance segmentation perform well, their output is somewhat unstructured and lacks musical significance. Therefore, a parallel step is necessary to achieve a more organised format. Given the challenges in accurately detecting thin, elongated objects such as staff lines, which are crucial for determining the pitch of a note, we have opted for a more efficient post-processing approach using traditional computer vision techniques that do not require training. In Figure \ref{fig:mask-output}, you can observe the results of running the Mask R-CNN model on a score image in the top image, and the same image with staff detection in the bottom Figure \ref{fig:mask-output-staff}, similarly for the object detection task shown in Figure \ref{fig:full-faster}.

\section{\uppercase{Conclusions}}
\label{sec:conclusion}

This study uses advanced neural network architectures to evaluate object detection and instance segmentation for OMR. Our findings confirm that instance segmentation, particularly using Mask R-CNN, offers capabilities in delineating detailed and accurate representations of individual musical symbols compared to traditional object detection methods. Mask R-CNN's detailed pixel-level segmentation capability makes it particularly adept at handling the complex visual compositions of sheet music where objects frequently overlap.

By comparing object detection and instance segmentation, we comprehensively evaluate their respective performances in OMR. This comparison is significant as it informs the selection of the most appropriate method depending on the specific requirements of a given task—whether broad localisation of symbols is sufficient or if precise delineation is necessary. Our findings contribute to a deeper understanding of how these techniques can be applied to optimise OMR systems.

One limitation of our approach is handling rare musical symbols, which are underrepresented in the training dataset. This could affect the generalisability of the model. Further research should focus on refining the detection of infrequent musical symbols and enhancing mask predictions for structurally complex objects. Prospective improvements could include optimising RPN anchor ratios and experimenting with larger training datasets to improve detection accuracy. 

This study highlights the potential of instance segmentation to advance OMR research, offering a more accurate and detailed method for recognising musical symbols in sheet music. These findings pave the way for further research and practical applications in musicology and digital archiving. Improved precision in symbol recognition directly impacts downstream tasks in knowledge discovery and information retrieval, enabling richer metadata generation for indexing and querying in music information retrieval systems. This facilitates sophisticated analysis, such as identifying patterns across compositions, discovering relationships between works, and conducting large-scale comparative studies. The detailed segmentation provided by Mask R-CNN supports more granular analysis, leading to deeper insights into musical structures and trends.

\section*{\uppercase{Acknowledgements}}

The authors acknowledge the support of the AI and Music CDT, funded by UKRI and EPSRC under grant agreement no. EP/S022694/1, and our industry partner Steinberg Media Technologies GmbH for their continuous support.



\bibliographystyle{apalike}
{\small
\bibliography{references}}



\end{document}